\documentclass[aps,prl,superscriptaddress,showpacs,twocolumn]{revtex4}

\usepackage{amsfonts}
\usepackage{amsmath}
\usepackage{amssymb}
\usepackage{graphicx}

\begin{document}

\title{Unexpected z-Direction Ising Antiferromagnetic Order in a frustrated Spin-1/2 $J_1-J_2$ \textit{XY} Model on the Honeycomb Lattice}

\author{Zhenyue Zhu}
\affiliation{Department of Physics and Astronomy, University of
California, Irvine, California 92697, USA}
\author{David A. Huse}
\affiliation{Physics Department, Princeton University, Princeton, New Jersey
08544, USA}
\author{Steven R. White}
\affiliation{Department of Physics and Astronomy, University of
California, Irvine, California 92697, USA}

\begin{abstract}
Using the density matrix renormalization group on wide
cylinders, we study the phase diagram of the spin-1/2 \textit{XY} model on
the honeycomb lattice, with first-neighbor ($J_1 = 1$) and
frustrating second-neighbor ($J_2>0$) interactions. For the
intermediate frustration regime $0.22\lesssim J_2\lesssim0.36$, we
find a surprising antiferromagnetic Ising phase, with ordered
moments pointing along the $z$ axis, despite the absence of any $S_z
S_z$ interactions in the Hamiltonian. Surrounding this phase as a
function of $J_2$ are antiferromagnetic phases with the moments
pointing in the $\textit{x-y}$ plane for small $J_2$ and a close competition
between an $\textit{x-y}$ plane magnetic collinear phase and a dimer phase
for large values of $J_2$. We do not find any spin-liquid phases in
this model.
\end{abstract}
\date{\today}
 \maketitle
\medskip

The past few years have seen a major resurgence in both experimental
and theoretical interest in quantum spin-liquid ground states
\cite{SL}. Much of the interest stems from strong evidence that
quantum spin liquids exist experimentally in several different
materials \cite{SL}. In the case of the kagome lattice material
herbertsmithite, ZnCu$_3$(OH)$_6$Cl$_2$, for example, mounting
experimental evidence for a spin-liquid low temperature phase
\cite{ex1,ex2,ex3,ex4,ex5} has coincided with recent strong
numerical evidence that the spin-1/2 Heisenberg kagome
antiferromagnet has a spin-liquid ground state \cite{kag1}, and that
this state has $Z_2$ topological order \cite{kag2,kag3}. The
numerical work has become possible through continued advances in
density matrix renormalization group (DMRG) techniques; these
methods can now be used to study frustrated spin Hamiltonians on
cylinders with widths up to 12 or 14 lattice spacings, which, when
combined with careful finite size analysis, can determine phases and
properties in the two-dimensional thermodynamic limit with good
confidence in many cases. There is great interest in understanding
the kagome spin liquid in more detail, and in finding other spin
liquids in simple realistic models.

Recently, Varney \emph{et al.} \cite{bose} studied the spin-1/2 \textit{XY}
model on the honeycomb lattice, with first-neighbor ($J_1=1$, $\langle
i,j\rangle$) and frustrating second-neighbor ($J_2 > 0$,
$\langle\langle i,j\rangle\rangle$) \textit{XY} interactions, with
Hamiltonian
\begin{equation}
H=J_1\sum_{\langle i,j\rangle}(S_i^+ S_j^- + H.c.)
+J_2\sum_{\langle\langle i,j\rangle\rangle}(S_i^+\ S_j^- + H.c.).
\end{equation}
Based on exact diagonalization (ED) of various small clusters, they
suggested that a particular spin-liquid ground state, a ``Bose
liquid,'' appears for $0.21\lesssim J_2 \lesssim 0.36$. Bose liquids
may have a singular surface in momentum space, similar to a Fermi
surface for a Fermi system, with gapless excitations and power-law
correlations \cite{bo1,bo2}, or they may be gapped and
incompressible \cite{bo3,mf}. This spin model is equivalent to
spinless hard-core bosons with first- and second-neighbor hopping
and zero off-site interactions. Our motivation for the present study
was to examine much larger lattices to see whether we could determine if
this system really has a spin-liquid ground state in some region of
the phase diagram in the thermodynamic limit.

\begin{figure}
\includegraphics*[width=8.5cm, angle=0]{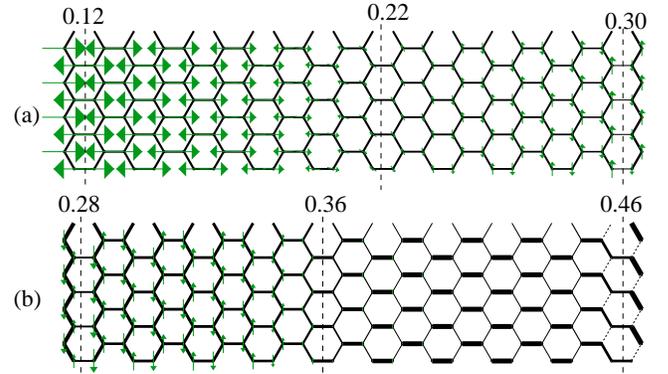}
\caption{(color online) (a) Local magnetic moments for an XC8
cylinder with $J_2$ varying along the length of the cylinder. The
dashed lines show the locations of particular $J_2$ values. In (a),
we show a cylinder with $J_2$ varying from 0.12 to 0.30. For
$J_2\sim 0.22$, the $xy$-plane Neel order rotates to $z$-direction
Ising order. In (b), $J_2$ is varied from 0.28 to 0.46. The second
phase transition point is located at $J_2\sim0.36$.}\label{phase}
\end{figure}

\begin{figure}
\includegraphics*[width=8cm, angle=0]{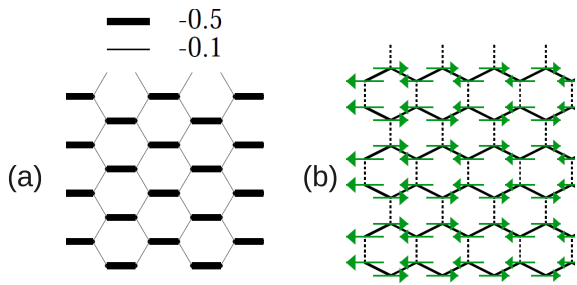}
\caption{(color online) (a) The ground state on the XC8 cylinder at
$J_2=0.5$, showing strong dimer correlations aligned horizontally.
We call this state a dimer state. (b) The ground state on YC6
cylinder at $J_2=0.5$. It has $xy$-plane collinear magnetic order,
with antiferromagnetic chains along the horizontal zigzag direction
together with ferromagnetic first-neighbor correlations between
these zigzag chains. }\label{phase3}
\end{figure}

We have performed numerous DMRG \cite{dmrg1,dmrg2,rev} calculations
on this model on long cylinders with circumferences up to 12 lattice
spacings.  The properties of the ground state are governed by the
ratio $J_2/J_1$. In all of our calculations, we take $J_1=1$ and $0\leq
J_2\leq 1$, thus antiferromagnetic interactions.  The cylinder
geometries we used in our DMRG calculations are adopted from Ref.
\cite{cyc}. For example, XC8 represents a cylinder where one set of
edges of each hexagon lie along the $x$ direction (which always
coincides with the cylinder axis), and there are 8 spins along the
circumferential zigzag columns, connected periodically (Fig.
\ref{phase}). The actual circumference (Euclidean distance) of XC8
is $C = 4\sqrt{3}$ lattice spacings. For the YC6 cylinder, one set
of edges of each hexagon lies along the $y$ (circumferential)
direction and there are 6 spins (in 3 pairs) along a straight
circumference of $C = 9$ lattice spacings [Fig. \ref{phase3}(b)].
For narrow cylinders like XC8, we are easily able to achieve a
truncation error of about $10^{-8}$ with M = 2400 states, which
determines the ground state essentially exactly. For YC8, our widest
cylinder with $C=12$, we need to keep M = 5800 to achieve a
truncation error of $10^{-6}$---still excellent accuracy. In all our
DMRG calculations, we keep enough states to make sure that the
truncation error is smaller than $10^{-6}$.

In the unfrustrated limit of $J_2=0$, the ground state has the
expected Neel order in the $xy$ plane. We find that this phase
extends to $J_2\sim 0.22$. In the interval $0.22\lesssim J_2\lesssim
0.36$, we find an antiferromagnetic phase that surprisingly has
staggered magnetization polarized along the $z$-direction in spin
space; we call this Ising antiferromagnetic order, to distinguish it
from Neel order in the $xy$ plane. Finally, for $J_2\gtrsim 0.36$,
we find that there is a close competition between a magnetically
ordered $xy$-plane collinear phase and a magnetically disordered
dimer phase. For example, on XC8 cylinders at $J_2=0.5$, the ground
state is a dimer state, with energy $E=-0.3227$; see Fig.
\ref{phase3}(a). A collinear state with an energy $1.2\%$ higher on
XC8 is metastable within DMRG. Similarly, on XC12 cylinders the
ground state is also a dimer state with energy only $0.25\%$ lower
than the collinear state. However, on XC10, YC4, and YC6 cylinders,
the ground state at $J_2=0.5$ is the collinear state, and the dimer
state is not even metastable; see Fig. \ref{phase3}(b) for an
illustration of the collinear ground state on the YC6 cylinder. The
collinear states on different cylinders have relatively small finite
size effects, with energy $E\cong -0.3189$. In this letter, we will
not try to resolve this close competition between these two states
for $J_2>0.36$ in the 2D limit; instead, we will focus on the
intermediate Ising phase regime.

From our DMRG calculations on large cylinders, we agree with Ref.
\cite{bose} about the rough locations of the two phase boundaries
and the properties of the phase for small $J_2$. However, in the
intermediate phase we find long-range Ising antiferromagnetic order,
which was not noticed in previous work on smaller systems. Thus we
conclude that this system does not have a spin-liquid phase. In
terms of bosons, this intermediate Ising phase has ``charge-density"
order of the bosons, with higher density on one sublattice than the
other.

In Fig. \ref{phase}, we present two cylinders to first give a quick
summary of the whole phase diagram. These are XC8 cylinders in which
$J_2$ is varied along the length of the cylinder, showing locally
the various phases. In Fig. \ref{phase}(a), $J_2$ varies from 0.12
to 0.30. At the $J_2=0.12$ left edge, a staggered field in the $xy$
plane was applied to ``pin" the Neel order. The ordered moments
rotate from the $x$ to the $z$ direction, indicating that there is a
phase transition between Neel and Ising order, at $J_2\sim0.22$. In
Fig. \ref{phase}(b), $J_2$ is varied from 0.28 to 0.46, with pinning
bonds at the $J_2=0.46$ right end to pin dimer order. The phase
transition from Ising to dimer order is visible at $J_2\sim0.36$. We
also applied these methods to other cylinders and find that the
values of $J_2$ at the estimated phase transitions change only
slightly between different width and orientation cylinders. Thus the
locations of these phase transitions show only small finite size
effects, which is consistent with our agreement with the small-size
ED results from Ref. \cite{bose}.

We have tested the stability of the Ising phase in several ways. For
example, one can measure the decay of the local staggered
magnetization away from an applied staggered field on an end of the
cylinder. For the Neel ordered phase (small $J_2$), when we apply
the pinning magnetic field along the $z$ direction, $|\langle
S_z\rangle|$ decays exponentially from the cylinder end [Fig. 3(a)].
To similarly test the Ising phase, we apply the pinning field along
the $x$ direction at the ends of an XC8 cylinder with $J_2=0.3$.  We
find that $|\langle S_x\rangle|$ decays exponentially with distance
from the cylinder end with a very short correlation length
$\xi_x=1.8$, but $|\langle S_z\rangle|$ rises from the end and
saturates in the center of cylinder (not shown). This provides solid
evidence that Ising order is very robust on this cylinder. As
another test, we have measured the correlation function $|\langle
S_i^+(0) S_j^-(x) \rangle|$ and find that its correlation length
decreases as a function of increasing $J_2$ for $J_2$ near 0.22, and
then increases rapidly for $J_2$ near 0.36. The minimum correlation
length is roughly $\xi\sim1.5$ at $J_2=0.3$ (not shown). This result
again confirms that $xy$-plane order is absent in the intermediate
phase.

\begin{figure}
\includegraphics*[width=8.5cm,angle=0]{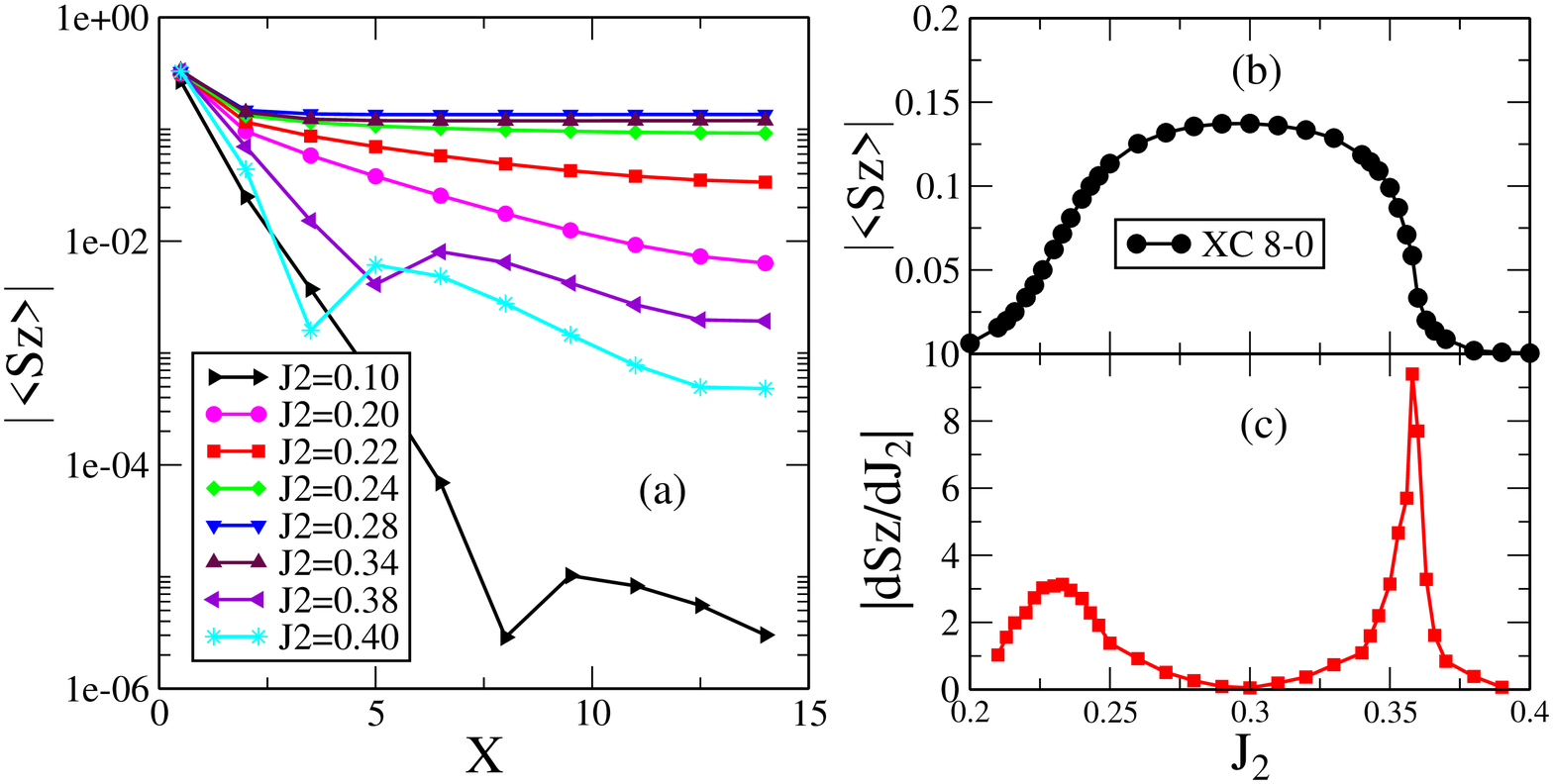}
\caption{(Color line) (a) Local magnetization $|\langle S_z\rangle|$
versus distance from the end of an XC8 cylinder for various $J_2$
values. (b) The magnetization at the cylinder center versus $J_2$.
(c) The derivative of the central $|\langle S_z\rangle|$ versus
$J_2$. The peaks of the derivative at $J_2=0.23$ and $0.36$ indicate
the two phase transition points. }\label{psz}
\end{figure}

In Fig. \ref{psz}, we apply a staggered field with $h_z=0.5$ at a
cylinder end to measure the decay of $|\langle S_z\rangle|$ with
distance from the end for various values of $J_2$. As shown in Fig.
\ref{psz}(a),  $|\langle S_z\rangle|$ decays exponentially within
both the Neel and the dimer phases, but the correlation length gets
longer and $|\langle S_z\rangle|$ becomes spatially uniform in the
cylinder center for the Ising ordered phase. In Fig. \ref{psz}(b),
we show the magnetization in the cylinder center versus $J_2$.  It
is clear from this plot that the intermediate Ising phase is a broad
regime, and from its derivative versus $J_2$, we determine the two
phase transition points at $\sim 0.23$ and 0.36, which approximately
match the phase transitions determined from Fig. \ref{phase}.  It is
interesting to note that $|\langle S_z\rangle|$ is almost
independent of $J_2$ for much of this intermediate Ising phase.  The
moment $|\langle S_z\rangle| \sim 0.14$ is strongly reduced from the
maximum ``classical'' value of 0.5.

The derivative of $|\langle S_z\rangle|$ shown in Fig. \ref{psz}(c)
shows markedly different behavior for the two transitions, with the
second transition being much sharper. A natural interpretation is
that the phase transition between Neel and Ising phases is
continuous, but the second phase transition point is first order.
To test this, we have performed calculations on cylinders with a
much narrower range of $J_2$ values along the length of the
cylinder, zooming in on the transitions. If the phase transition is
first order, we expect that the phase transition region should
remain narrow as we zoom in. For a continuous phase transition, the
phase transition region should broaden as we zoom in. Varying the
gradient of $J_2$ by a factor of 5, we do find that the Neel-Ising
phase transition region broadens, but the second phase transition
region stays narrow.  Thus, it does appear that the former is
continuous, and the latter is first order. However, any conclusions
about the second transition are tentative, because of the close
competition between the dimer and collinear phases for $J_2>0.36$.

\begin{figure}
\includegraphics*[width=8.5cm,angle=0]{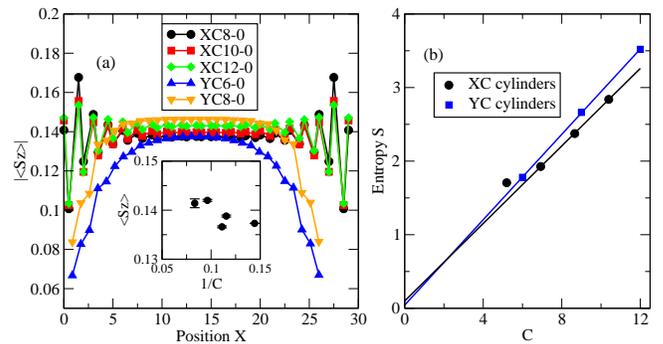}
\caption{(color online) (a) The absolute value of the local
magnetization $|\langle S_z\rangle|$ for various XC and YC cylinders
versus distance along the cylinder. The inset shows the extrapolated
magnetization (extrapolated versus the truncation error) at the
cylinder center, with error bars, versus the inverse of cylinder
circumference $C$. (b) The entanglement entropy versus circumference
for XC and YC cylinders in the intermediate Ising ordered phase at
$J_2=0.3$. The intercepts are consistent with zero.}\label{mag-s}
\end{figure}

To make sure that the Ising order is not a finite size effect, we
have studied the $J_2=0.3$ system for cylinders with various widths.
Figure \ref{mag-s}(a) shows $|\langle S_z\rangle|$ as a function of
$x$ for XC and YC cylinders. In the inset, we plot the extrapolated
magnetization at the cylinder center versus the inverse of the cylinder
circumference.  For these cylinders, the staggered magnetization is
nearly constant with circumference, taking a value of about
$0.135-0.142$. Thus, we believe that $|\langle
S_z\rangle|\sim0.14$ in the 2D limit for $J_2=0.3$.  If anything,
the staggered magnetization increases with increasing $C$, so this
should be viewed as a lower bound on the value in the 2D limit.

We also measured the entanglement entropy for various cylinder sizes
and extrapolated to see if there is a possible topological
entanglement entropy contribution ($\gamma$) \cite{ent1,ent2} see Fig.
\ref{mag-s}(b). 
Entanglement entropy area law states that for a gapped phase 
$S\sim a L+\gamma+O(1/L)$, with L its boundary length\cite{ea}. 
For a $Z_2$ spin liquid, one would expect
$\gamma=-\ln 2$.  We find $\gamma \sim 0.09$ for XC cylinders and
$\gamma \sim 0.04$ for YC cylinders, values consistent with zero, as
expected for a nontopologically ordered state. We expect that if we
could include larger cylinders, the resulting data would extrapolate
to $\gamma = 0$.

\begin{figure}
\includegraphics*[width=8.5cm,angle=0]{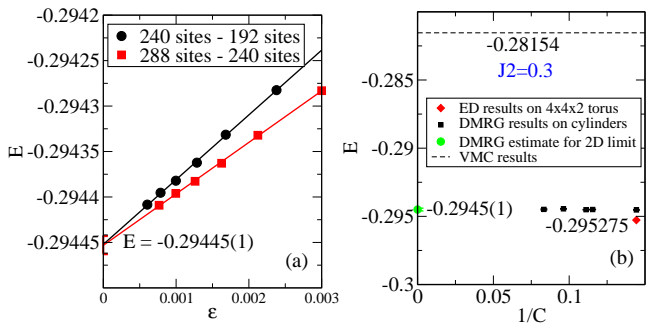}
\caption{(color online) (a) The extrapolation of the ground state
energy per spin for $J_2=0.3$ versus truncation error for the XC12
cylinder. The black curve is the energy per spin from subtracting
the energies of two XC12 cylinders with lengths $L_x =20 $ and
$L_x=16$. The red curve is from subtracting two cylinders with $L_x
=24 $ and $L_x=20$. These two subtractions extrapolate to the same
energy per spin of $-0.29445(1)$. (b) The ground state energy per
spin for $J_2=0.3$ versus the inverse of cylinder circumference from
our DMRG calculations, compared with the variational Monte Carlo (VMC)
result from Table III in the supplemental material of Ref.
\cite{vmc}, and exact diagonalization \cite{vmc}.}\label{ener}
\end{figure}

Very recently, a variational Monte Carlo study of this model has
appeared \cite{vmc}. In Ref. \cite{vmc}, a variational spin-liquid
wave function is constructed by decomposing the boson operators into
a pair of fermions with a long-range Jastrow factor, with Gutzwiller
projection enforcing single occupancy. At $J_2=0.3$, the lowest
energy for such a state had energy per spin $E=-0.28154$, which is
$\sim 4\%$ higher than ED of the
$4\times4\times2$ torus ($E=-0.295275$) \cite{vmc}. Although this
might appear to be a small difference in energy, for competing
phases in geometrically frustrated spin-1/2 models near spin
liquids, this is actually a very large energy difference.  For
example, the spin-1/2 kagome antiferromagnet (with $J_2=0$) has an
energy difference of only about $1\%$ between the (metastable)
honeycomb valence bond crystal and the spin-liquid ground state
\cite{kag1}. In our DMRG calculations, the energy per spin for a
specific cylinder geometry can be calculated by subtracting two
cylinders with the same width but different lengths \cite{rev}. When
the cylinder is long enough, this method gives the energy per spin
in the cylinder center, with minimal edge effects. We show in Fig.
\ref{ener}(a) that the energy per spin from subtracting two
different pairs of cylinders gives precisely the same energy for the
XC12 cylinder. Thus, we find that the ground state energy per spin is
-0.29445(1) for an infinitely long XC12 cylinder at $J_2=0.3$. In
Fig. \ref{ener}(b), we compare our DMRG results for the ground state
energy on various cylinders at $J_2 = 0.3$. For the cylinders we
study, the DMRG energies have quite small finite size effects. We
estimate that the ground state energy is $E=-0.2945(1)$ in the 2D
limit. The small-size ED result is only slightly ($\sim 0.26\%$)
lower in energy, due to its finite size effects.  The state DMRG
finds has antiferromagnetic Ising order with the spin moments
ordered in the $z$ direction. This ordered ground state has much
lower energy than the variational spin-liquid state.

We have not been able to find a simple analytical argument or
calculation that gives an intuitive picture for this Ising ordered
state.  However, viewing the system as hard-core bosons at
half filling provides an additional perspective. The Hamiltonian can
be mapped straightforwardly and exactly into a hard-core boson model
with first-neighbor hopping $t_1 = J_1/4$ and second-neighbor
hopping $t_2 = J_2/4$, since $S^{\dagger}=b^{\dagger}/2$,
$S^z=b^{\dagger}b-0.5$. The Ising order would appear as a charge
density wave (CDW) order, where the density is higher on sublattice
\textit{A} ($n_A\sim 0.64$) than on sublattice \textit{B} ($n_B=1-n_A\sim 0.36$).
Although there are only hopping terms in this hard-core boson
Hamiltonian, the hard-core constraint (one boson per lattice site)
is an on-site interaction.  One could imagine that this on-site
interaction renormalizes in some way to produce a first-neighbor
density-density interaction, which could produce the CDW. This
system is the first that we are aware of where a CDW is produced
only from the combination of frustrated hopping and a hard-core
constraint.

In summary, we have studied the $J_1-J_2$ antiferromagnetic spin-1/2
\textit{XY} model on the honeycomb lattice on various cylinders with DMRG.
Instead of a spin-liquid ground state in the intermediate phase
regime for $0.22<J_2/J_1<0.36$, we find an Ising ordered phase with
a staggered magnetization along the $z$ direction that does not show
any strong finite size effects. This Ising order is thus robust on
various cylinder circumferences.  We obtain a ground state energy
much lower than proposed spin-liquid states, and a vanishing
topological entanglement entropy.  Thinking about this in terms of
the spin model, it is somewhat puzzling to understand why this phase
appears, since there are no $S^z_iS^z_j$ interaction terms in the
spin Hamiltonian. Describing the system instead as hard-core bosons
with frustrated hopping, this Ising phase is then a Mott insulator
with one boson per two-site unit cell, and the Ising order is then
CDW order that breaks the $Z_2$ sublattice symmetry of the unit
cell. The on-site hard-core interaction must induce a first-neighbor
repulsion that stabilizes this CDW order. Thus, although this model
unfortunately does not appear to exhibit a spin-liquid ground state,
it exhibits this somewhat surprising CDW $Z_2$ ordered
phase.

\begin{acknowledgments}

We thank Hongcheng Jiang for early collaboration on this work.  We
would also like to thank Sasha Chernyshev, Leonid Glazman, Andreas
Laeuchli, Sid Parameswaran, Marcos Rigol, Victor Galitski, Leon
Balents, Miles Stoudenmire, and Simeng Yan for many helpful
discussions.  This work was supported by NSF Grant No. DMR-1161348 (Z.Z.,
S.R.W.), NSF MRSEC Grant No. DMR-0819860 (D.A.H.), and the DARPA OLE program
(D.A.H.).

\end{acknowledgments}

\end{document}